\documentstyle[aasms4,amstex,amsfonts,epsfig,rotating,float,12pt]{article}

\def\refindent{\par\noindent\hangindent=3pc\hangafter=1 }
\def\aa#1#2#3{\refindent#1, A\&A, #2, #3}

\def\apj#1#2#3{\refindent#1, {\it ApJ}, {\bf#2}, #3.}
\def\apjlett#1#2#3{\refindent#1, {\it ApJ (Letters)}, {\bf #2}, #3.}

\def\>{$>$}
\def\<{$<$}

\def\simlt{\lower.5ex\hbox{$\; \buildrel < \over \sim \;$}}
\def\simgt{\lower.5ex\hbox{$\; \buildrel > \over \sim \;$}}
\def\sqr#1#2{{\vcenter{\hrule height.#2pt
      \hbox{\vrule width.#2pt height#1pt \kern#1pt
         \vrule width.#2pt}
      \hrule height.#2pt}}}

\begin{document}

\centerline{Submitted to the Astrophysical Journal (Letters)}
\centerline{Revised June 8, 1998}
\bigskip
\title{A Model of the EGRET Source at the Galactic Center:\\
       Inverse Compton Scattering Within Sgr A East and its Halo}

\author{Fulvio Melia\altaffilmark{1}$^{*\dag}$, 
Farhad Yusef-Zadeh$^{\ddag}$
and Marco Fatuzzo$^*$}
\affil{$^*$Physics Department, The University of Arizona, Tucson, AZ 85721}
\affil{$^{\ddag}$Department of Physics and Astronomy, Northwestern
University, Evanston, IL 60208}
\affil{$^{\dag}$Steward Observatory, The University of Arizona, Tucson, AZ 85721}





\altaffiltext{1}{Presidential Young Investigator.}


\begin{abstract}
Continuum low-frequency radio observations of the Galactic Center reveal 
the presence of two prominent radio sources, Sgr A East and its surrounding 
Halo, containing non-thermal particle distributions with power-law indices 
$\sim 2.5-3.3$ and $\sim 2.4$, respectively.  The central $1-2$ pc region 
is also a source of intense (stellar) UV and (dust-reprocessed) far-IR 
radiation that bathes these extended synchrotron-emitting structures.  A 
recent detection of $\gamma$-rays (2EGJ1746-2852) from within $\sim 1^o$ 
of the Galactic Center by EGRET onboard the Compton GRO shows that the 
emission from this environment extends to very high energies.

We suggest that inverse Compton scatterings between the power-law electrons 
inferred from the radio properties of Sgr A East and its Halo, and the UV and 
IR photons from the nucleus, may account for the possibly diffuse $\gamma$-ray 
source as well.  We show that both particle distributions may be contributing
to the $\gamma$-ray emission, though their relevant strength depends on
the actual physical properties (such as the magnetic field intensity)
in each source. If this picture is correct, the high-energy source at the Galactic
Center is extended over several arcminutes, which can be tested with the
next generation of $\gamma$-ray and hard X-ray missions.
\end{abstract}




%

\section{Introduction}
In 1992, EGRET on board the Compton GRO identified a central
($< 1^o$) $\sim 30$ MeV - $10$ GeV continuum source with luminosity 
$\approx 2\times 10^{37}$ ergs s$^{-1}$ (\cite{mayer98}). 
Its spectrum can be represented as a broken hard power-law 
with spectral indices $\alpha=-1.3\pm 0.03$ and $-3.1\pm
0.2$ ($S=S_0\,E^\alpha$), with a cutoff between $4-10$ GeV.  This 
EGRET $\gamma$-ray source (2EGJ1746-2852) appears to be centered 
at $l\approx 0.2^o$, but a zero (or even a negative) longitude 
cannot be ruled out completely.  

The $\gamma$-ray flux does not appear to be variable down to the 
instrument sensitivity (roughly a $20\%$ amplitude), which has led some 
to postulate that the observed $\gamma$-rays are produced by diffuse 
sources, either within the so-called Arc of non-thermal filaments 
(\cite{po97}), or as a result of the explosive event forming the large
supernova-like remnant Sgr A East (\cite{yz97}).  (A schematic
diagram of the morphology of the central parsecs is shown in Fig. 
1 below.)  Markoff, Melia \& Sarcevic (1997) also considered in detail 
a possible black hole 
origin for the $\gamma$-rays under the assumption that the ultimate 
source of power for this high-energy emission may be accretion 
onto the central engine.  They concluded that it is not yet 
possible to rule out Sgr A* (which appears to be coincident with
the central dark mass concentration) as one of the possible sources 
for this radiation, and that the expected spectrum is a good 
match to the observations.  The lack of variability larger than 
$\sim 20\%$ in the high-energy flux would then be consistent 
with the maximum amplitude of the turbulent cell fluctuations seen 
in three-dimensional hydrodynamical simulations of accretion onto 
Sgr A* (\cite{ruf94}; \cite{cm97}).  It appears that a true 
test of Sgr A* as the source for the EGRET emission would be the 
detection (or non-detection) of variability with an 
amplitude significantly smaller than this.  

To answer the question of whether or not 2EGJ1746-2852 is coincident 
with Sgr A*, it is essential to fully understand the alternative 
contributions to the high-energy flux from the Galactic Center.  The unique 
environment in this region facilitates the co-existence of thermal
and non-thermal particles, which can lead to interactions that
produce a substantial diffuse Compton upscattering emissivity.  There
is now considerable evidence that the radio spectrum of Sgr A East
and the Halo is likely synchrotron radiation by
nonthermal particles at high energy (Pedlar, et al. 1989).  
However, this region is also bathed with an intense ultraviolet 
(UV) and infrared (IR) photon field from the central $1-2$ parsecs
and these particles must therefore be subjected to numerous Compton
scattering events.  Our focus in this {\it Letter} is to see 
whether the properties of this relativistic electron distribution, 
inferred from their observed radio characteristics, also make them
a viable source for the $\gamma$-rays detected by EGRET.  This is
particularly important in view of the fact that it may be possible
to distinguish between Sgr A* and an extended $\gamma$-ray source
with high-resolution $\gamma$-ray (or hard X-ray) imaging.  For example, 
the proposed balloon flight instrument UNEX (Rothschild 1998) may
have sufficient sensitivity to image the hard X-ray counterpart to 
2EGJ1746-2852.

\section{Sgr A East, the Halo and the Galactic Center Radiation Field}
Radio continuum observations of the Galactic center show a prominent 
nonthermal radio continuum shell-like structure, Sgr A East, as well as 
thermal ionized gas, known as Sgr A West, orbiting Sgr A$^*$ (Ekers, et al. 
1983; Pedlar, et al. 1989; Serabyn et al. 1991).  The latter 
two are themselves surrounded by a torus of 
dust and molecular gas known as the Circumnuclear Disk (CND).
Figure 1 shows a schematic diagram of Sgr A East, its Halo, and
their location relative to the black hole candidate Sgr A*,
centered within the CND. Low-frequency continuum 
observations show a deep depression in the brightness of the Sgr 
A East shell at the position of Sgr A West, which results
from free-free absorption of the radiation from Sgr A East by 
the thermal gas in Sgr A West.  Sgr A East must therefore
lie behind Sgr A West (Yusef-Zadeh \& Morris 1987; Pedlar et al. 1989). 
The exact distance of Sgr A East behind Sgr A West is not known, 
but a number of arguments suggest that it is located very close to 
the Galactic Center (e.g., G\"usten \& Downes 1980; Goss et al. 1989). 

On a larger scale, there is a diffuse $7^{\prime}-10^{\prime}$ Halo of 
nonthermal continuum emission surrounding the oval-shaped radio 
structure Sgr A East. Assuming a power-law distribution of 
relativistic particles, the energy spectrum of the relativistic 
electrons within the shell and the Halo are estimated to be $\sim 2.5-3.3$ 
and $\sim 2.4$, respectively (Pedlar et al. 1989). The Halo may be a 
secondary manifestation of the explosion that produced Sgr A East.
However, the fact that the particle spectral index is steeper in
the latter suggests that significant cooling of its relativistic
particles may already have taken place which may not be consistent
with a model in which the cosmic-ray electrons leak through
the shell and produce the extended Halo radio emission.  Thus,
the Halo may be unrelated to the creation of Sgr A East, as
Pedlar et al. (1989) have suggested.  
It may instead be associated with continued activity at the Galactic 
Center, possibly from the expansion of relativistic particles
that are not confined by Sgr A*.  This may also be taken as indirect
evidence that the Halo, unlike Sgr A East, may be centered on
Sgr A* (see Fig. 1).  In either
case, what is of interest to us here is the indication from 
radio observations of the presence of these power-law particle 
distributions in the extended region surrounding Sgr A*.  
The Compton spectrum from a lepton distribution with index $p
\equiv 2.4-3.3$ is expected to have a spectral index $\alpha\sim 
(1+p)/2\approx 1.7-2.2$, close to that of 2EGJ1746-2852.    

The optical depth toward Sgr A East and the Halo at low frequencies 
led Pedlar et al. (1989) to consider a mixture of both thermal and 
nonthermal gas, though displaced to the front side of Sgr A East. 
Pedlar et al. (1989) also showed evidence that the nonthermal
emission from the Halo is located in front of the thermal gas in
Sgr A West. The schematic diagram in Figure 1 depicts a geometry in which 
the Sgr A East shell lies close to, but behind, the Galactic Center 
whereas the diffuse Sgr A East Halo surrounds the Galactic 
Center and the shell. 

\begin{figure}[H] 
\centerline{\begin{turn}{0}\epsfig{file=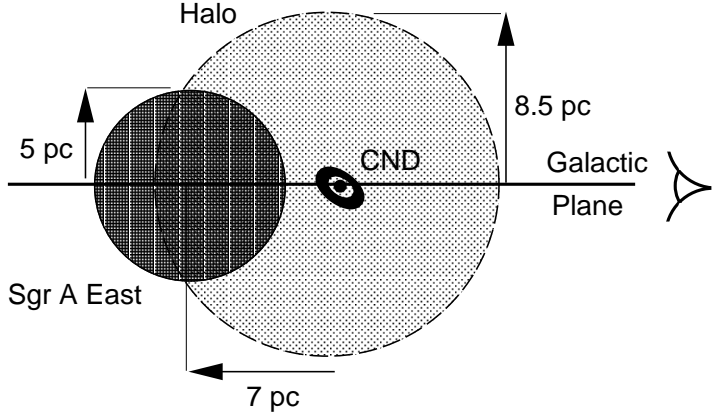,width=4.0in}\end{turn}}
\vspace{10pt}
\caption{Schematic diagram showing the relative positions and
sizes of the Halo and Sgr A East relative to Sgr A*, which is
shown here as a point centered within the CND. The thermal 3-arm
spiral radio source Sgr A West is also contained within the CND.}
\label{fig1}
\end{figure}

At $\lambda 20$ cm, Sgr A East and the Halo are among
the brightest radio sources in the sky, with integrated flux
densities of $222$ and $247$ Jy, respectively (Pedlar, et al. 1989).
Their average brightness is about 900 and 350 mJy per 12" beam, 
respectively, and in order to fit the
radio spectrum, the maximum particle Lorentz factor in these
sources should be $\gamma_{max}\sim 2\times 10^5$. Thus, with
a minimum Lorentz factor $\gamma_{min}\sim 6,000$ (see discussion
in the following section), the total number of radiating electrons
is  
\begin{equation}\label{Nhalo}
N_0(\hbox{Halo})\approx 1.0\times 10^{52}\;\left({10^{-5}\;\hbox{G}\over
B\sin\theta}\right)^{1.7}\;,
\end{equation}
and
\begin{equation}\label{Nsgrae}
N_0(\hbox{Sgr A East Shell})\approx 4.2\times 10^{52}\;\left({10^{-5}\;\hbox{G}\over
B\sin\theta}\right)^{2.15}\;,
\end{equation}
where $B$ is the magnetic field and $\theta$ is the pitch
angle. The Halo particles are assumed to be distributed uniformly throughout
its volume.  On the other hand, the Synchrotron emission from Sgr A East is
concentrated within a shell with thickness $d\sim 1$ pc (e.g., Pedlar, et al. 1989).  
However, the gyration radius $a_{gyr}$ within this region with $B\sim 10^{-5}$ G is
$\sim 3\times 10^{13}$ cm for the most energetic particles ($\gamma_{max}=
2\times 10^5$), and so the diffusion time out of the shell is expected to
be $\sim (d/c)(d/a_{gyr})\approx 3\times 10^5$ years, much longer than the
$\tau_{age}\sim 10,000$ year lifetime of the remnant. So most of the Compton
scatterings associated with Sgr A East are expected to occur within its shell. 

These relativistic particles are immersed in an intense source of
UV and IR radiation from the central $1-2$ parsecs.  The ring of
molecular gas (also known as the Circumnuclear Disk, or CND) 
is rotating, and is heated by a centrally concentrated source of
UV radiation (predominantly the IRS 16 cluster of bright, blue stars).  
The CND encloses a central concentration of dark matter, which is believed to
be a $\sim$ 2.6$\times$10$^{6}$ solar mass black hole
(e.g., Haller et al. 1996; Genzel et al. 1996).
The CND is a powerful source ($\approx 10^7\:L_\odot$) of 
mid to far-infrared continuum emission with a dust temperature of 
$\approx 100$ K (e.g., Telesco et al. 1996; Davidson et al. 1992). 
This radiation is due to reprocessing by warm dust 
that has absorbed the same power in the UV (Becklin, Gatley and Werner 1982;
Davidson et al. 1992). Models of the photodissociation regions in the CND require
an incident dissociating flux ($6$ eV $< h\nu < 13.6$ eV) of $10^2$--$10^3$
erg cm$^{-2}$ s$^{-1}$ (Wolfire, Tielens \& Hollenbach, 1990), implying 
a total UV luminosity of about $2\times 10^7\;L_\odot$, consistent with the 
radio continuum emission from Sgr A West (Genzel et al. 1985).  This
intensity is also suggested by the detection of radio continuum
emission from the outer envelope of IRS 7, a cool supergiant 
being photoionized by the UV radiation field (e.g., Serabyn et al. 1991; 
Yusef-Zadeh and Melia 1992), and is
consistent with the inferred ionizing flux in Sgr A West, corresponding to a
centrally concentrated source of $2\times 10^{50}$ ionizing photons
per second (Lacy, et al. 1982; Ekers, et al. 1983; Mezger and Wink 1986).

\section{Inverse Compton Scattering within Sgr A East and the Halo}
The dominant cooling mechanism for the relativistic electrons as they diffuse
throughout the Sgr A East and Halo regions is inverse Compton scatterings with 
the Galactic Center stellar UV photons and the
reprocessed IR photons from the CND.  This radiation field has a specific
photon number density per solid angle $n_{ph}^{tot}(\varepsilon)\equiv
n_{ph}^{UV}(\varepsilon)+n_{ph}^{IR}(\varepsilon)$, where 
$n_{ph}^{UV}(\varepsilon) = N_0^{UV}(2\varepsilon^2/h^3 c^3)
(\exp\{\varepsilon/kT^{UV}\}-1)^{-1}$, and $n_{ph}^{IR}(\varepsilon) = 
N_0^{IR}(2\varepsilon^3/h^3 c^3)(\exp\{\varepsilon/kT^{IR}\}-1)^{-1}$.
Here, $\varepsilon$ is the lab-frame photon energy and $T^{UV}$ and
$T^{IR}$ are, respectively, the temperature (assumed to be $30,000$ K)
of the stellar UV component and of the reprocessed CND radiation, which
is assumed to peak at $50\mu$m, corresponding to a characteristic temperature
$T^{IR}\approx 100$ K.  Note that these expressions take into account the
energy dependence of the efficiency factor for dust emission, which leads
to a modified blackbody form for the dust spectrum.  The normalization 
constants $N_0^{UV}$ and $N_0^{IR}$ incorporate the dilution in photon 
number density as the radiation propagates outwards from the central 
core.  For the UV radiation, this is calculated assuming that the radiation 
emanates from a sphere of radius $\approx 1$ pc, whereas
for the IR radiation, we assume a total luminosity of $10^7\,L_\odot$
from a disk with radius $\approx 2$ pc. 
 
In the following expressions, primed quantities denote values in the electron 
rest frame, whereas unprimed parameters pertain to the (stationary) lab frame.  
An electron moving with Lorentz factor $\gamma$ 
through this field scatters $dN$ photons to energies between 
$\varepsilon_s$ and $\varepsilon_s + d\varepsilon_s$ and solid angles between 
$\mu_s \phi_s$ and $[\mu_s + d\mu_s][\phi_s + d\phi_s]$ at a rate (per energy 
per solid angle)
\begin{equation}\label{dN}
{dN\over dt d\varepsilon_s d\mu_s d\phi_s} = \int_{ph} d\varepsilon 
\;\int\;d\mu\,d\phi\; n_{ph}^{tot}(\varepsilon,\Omega)\;
\left({d\sigma_{KN}\over d\mu_s' d\phi_s'
d\varepsilon_s'}\right) \;{c (1-\beta\mu)\over \gamma(1-\beta\mu_s)}\;,
\end{equation}
where $\beta = (1-\gamma^{-2})^{-1/2}$, $\mu$ is the cosine of $\theta$
relative to the electron's direction of motion, and $\phi$ is the azimuthal
angle that completes the integration over all solid angles.
The differential Klein-Nishina cross-section ${d\sigma_{KN}/ d\mu_s' d\phi_s'
d\varepsilon_s'}$ is evaluated in the electron rest frame.
Using the general expressions relating the lab and rest frame energies  ($\varepsilon' = 
\varepsilon\gamma[1-\beta\mu]$) and angles ($\mu' = [\mu-\beta]/[1-\beta\mu]; 
\phi'=\phi$), one finds the relation 
$d\varepsilon_s d\mu_s d\phi_s / d\varepsilon_s' d\mu_s' d\phi_s' = 
\gamma (1-\beta\mu_s)$, thereby allowing Equation (3) to be  easily 
integrated over all scattered photon energies and solid angles to yield 
the single electron scattering rate.  

The inverse Compton (X-ray and $\gamma$-ray) emissivity can be determined by 
integrating Equation (3) over the entire scattering electron population. For
simplicity, we assume that the electron distribution
is locally isotropic, which then also implies that the upscattered radiation
field is emitted isotropically from within a volume $V\sim 250$ pc$^3$
in the case of the Sgr A East shell and $\sim 2,500$ pc$^3$ for the Halo,
and corresponding surface area $4\pi R^2$, where $R\approx 5$ pc for
the former and $\approx 8.5$ pc for the latter (see Fig. 1). 
Thus, the rate at which photons are detected by an observer at a distance 
$D$ is given by the expression
\begin{equation}\label{dNtot}
{dN_{obs}\over dt d\varepsilon_s dA} = {1\over 2D^2}\; 
\int_{V}d^3x\;\int_{\gamma_{min}}^{\gamma_{max}}\; d\gamma \;\int_{-1}^1 d\mu_s\;
n_e(\gamma)\; {dN\over dt d\varepsilon_s d\mu_s d\phi_s}\;,
\end{equation}
where azimuthal symmetry has been invoked to eliminate the need
to average over $\phi_s$. Note that the integral over $\phi$
is still within ${dN/ dt d\varepsilon_s d\mu_s d\phi_s}$.
This expression also uses the relationship between the 
scattered photon solid angle $d\mu_s \;d\phi_s$ and the detector area 
$dA$ ($dA = D^2 d\mu_s d\phi_s$).

\section{Results and Discussion}
The main results of our calculations are summarized in Figures
2 and 3, which show, respectively, the cumulative spectra from
Sgr A East and the Halo. Both sources appear to account well for 
the high-energy spectrum of the Galactic center, though their
actual $\gamma$-ray fluxes depend on the magnetic field intensity
within each of the structures.  The value of $B$ assumed for
both of these figures is $1.8\times 10^{-5}$ G, which then fixes
the relativistic particle numbers quoted there (see also eqs. 1 and 2).

\begin{figure}[H] 
\centerline{\begin{turn}{0}\epsfig{file=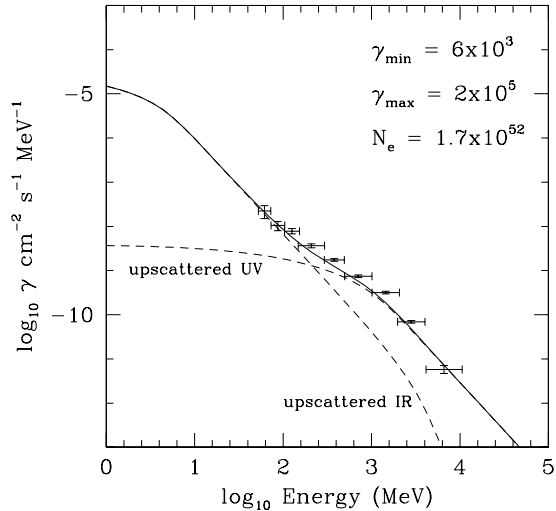,width=3.3in}\end{turn}}
\vspace{10pt}
\caption{Combined spectrum from inverse Compton scattering within Sgr 
A East.  The components shown here are: the upscattered IR, and the 
upscattered UV.  The cumulative spectrum is shown as a thin solid curve.  
The EGRET data are from Mayer-Hasselwander, et al. (1998).}
\label{fig2}
\end{figure}

The spectral turnover below $\sim 1$ GeV is difficult to produce with 
Compton scatterings without a low-energy cutoff in the particle
distribution.  Unlike the situation where the $\gamma$-rays
result from pion decays, in which this turnover is associated with 
the pion rest mass (Markoff, Melia \& Sarcevic 1997), there is no
natural characteristic cutoff energy here.  To match the data,
we have adopted a minimum Lorentz factor $\gamma_{min}\approx
6,000$, but we do not yet have a compelling argument for this value, 
though we can offer the following suggestion.  If the protons and electrons
continue to interact after they leave the shock acceleration region,
either Coulomb scatterings or a charge-separation electric field
can gradually shift the overall electron distribution
to a higher Lorentz factor due to the mass differential between the
two sets of particles.  If the electrons and protons are energized
more or less equally, then in a neutral plasma the lepton
$\gamma_{min}$ must be much greater than $1$.
For a proton index $\alpha_p$ and an electron index $\alpha_e$,
it is then evident that $\gamma_{min}\approx [(\alpha_p-1)/(\alpha_p-2)]
\times [(\alpha_e-2)/(\alpha_e-1)]\times (m_p/m_e)$.  In this context,
a $\gamma_{min}\approx 6,000$ for the electrons may therefore reflect 
the difference in particle mass and the relativistic distribution indices.
For this to work, the energy equilibration would have to occur {\it in situ},
perhaps due to a uniform acceleration of the relativistic electrons
by a charge separation-induced electric field, as mentioned previously.  
Clearly, if either Sgr A East and/or the Halo turn out to be the source 
of $\gamma$-rays, this explanation (or a viable alternative) must be 
developed more fully.

\begin{figure}[H] 
\centerline{\begin{turn}{0}\epsfig{file=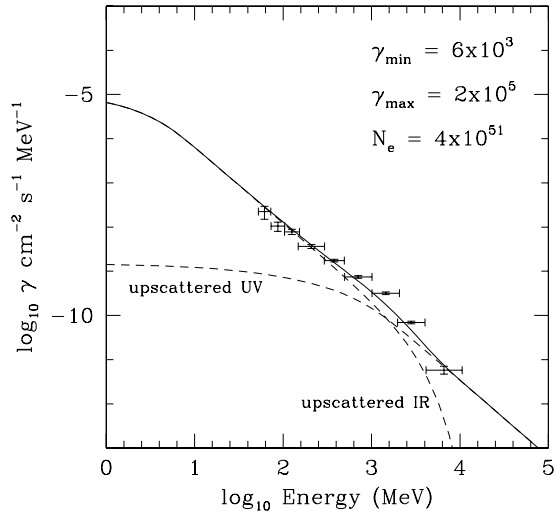,width=3.3in}\end{turn}}
\vspace{10pt}
\caption{Same as Fig. 2, except here for the Halo. This fit assumes
the same value of $B$ ($\sim 1.8\times 10^{-5}$ G) required to fit
the spectrum with Sgr A East's emissivity, which then gives a Halo
relativistic particle number $N_e=4\times 10^{51}$ as indicated.}
\label{fig3}
\end{figure}

If the $\gamma$-rays detected by EGRET are indeed the upscattered
UV photons from the Galactic center, it seems inevitable to us
that the corresponding IR photons from the CND should result in 
a significant upscattered intensity at intermediate (i.e.,
$\sim 10-100$ keV) energies.  This flux density ($\sim 10^{-5}$
photons cm$^{-2}$ s$^{-1}$ MeV$^{-1}$, or possibly higher if
$\gamma_{min}<6,000$) may be above the sensitivity limit (which is
$\sim 10^{-7}$ photons cm$^{-2}$ s$^{-1}$ MeV$^{-1}$ for a point
source) of UNEX, a proposed balloon flight instrument (Rothschild
1998).  With its expected spatial resolution of several arcmin or 
less, this experiment should therefore have little trouble imaging 
the hard X-ray counterpart to 2EGJ1746-2852, if this source is 
extended and is associated with either Sgr A East and/or the Halo. 

\section{Acknowledgments}
This work was supported by NASA grant NAGW-2518. We are grateful to
the anonymous referee, whose comments have led to a significant
improvement of our paper. 
%
%
%
{}


\begin{thebibliography}{}

\bibitem[Becklin, Gatley \& Werner 1982]{beck82}\apj{Becklin, E.E., Gatley, 
I. \& Werner, M.W. 1982}{258}{134}
\bibitem[Coker \& Melia 1997]{cm97}\apjlett{Coker, R.F. \& Melia, F. 1997}{488L}{149}.
\bibitem[Davidson et al. 1992]{dav92}\apj{Davidson, J.A., Werner, M.W., Wu, X.,
Lester, D.F., Harvey, P.M., Joy, M. \& Morris, M. 1992}{387}{189}
\bibitem[Ekers, et al. 1983]{ek83}\aa{Ekers, R.D., Van Gorkom, J.H.,
Schwarz, U.J. \& Goss, W.M. 1983}{122}{143}
\bibitem[Genzel, et al. 1985]{gen85}\apj{Genzel, R., Crawford, M.K., Townes, C.H.
\& Watson, D.M. 1985}{297}{766}
\bibitem[Genzel, et al. 1996]{gen96}\apj{Genzel, R., et al. 1996}{472}{153}
\bibitem[Goss, et al. 1989]{goss89}\refindent Goss, M. et al. 1989, The center                   
of the Galaxy, IAU 136 ed. M. Morris, p345.
\bibitem[G\"usten \& Downes 1980]{gd80}\aa{G\"usten, R. \& Downes, D. 1980}{87}{6}
\bibitem[Haller, et al. 1996]{hall96}\apj{Haller, J.W.,
Rieke, M.J., Rieke, G.H., Tamblyn, P., Close, L. \& Melia, F.
1996}{456}{194}
\bibitem[Lacy, Townes \& Hollenback 1982]{lac82}\apj{Lacy, J.H., Townes, C.H.
\& Hollenbach, D.J. 1982}{262}{120}
\bibitem[Markoff, Melia \& Sarcevic 1997]{mar97}\apjlett
{Markoff, S., Melia, F. \& Sarcevic, I. 1997}{489L}{47} (Paper I)
\bibitem[Mayer-Hasselwander, et al. 1998]{mayer98}\refindent Mayer-Hasselwander, 
H.A., et al. 1998, A\&A, in press.
\bibitem[Mezger \& Wink]{mez86}\aa{Mezger, P.G. \& Wink, J.E. 1986}{157}{252}
\bibitem[Pedlar, et al. 1989]{ped89}\apj{Pedlar, A., et al. 1989}
{342}{769}
\bibitem[Pohl 1997]{po97}\aa{Pohl, M. 1997}{317}{441}.
\bibitem[Rothschild 1998]{roth98}\refindent Rothschild, R. 1998, private communication.
\bibitem[Ruffert \& Melia 1994]{ruf94}\aa{Ruffert, M. \& Melia, F. 1994}
{288}{L29}.
\bibitem[Serabyn, Lacy \& Achtermann 1991]{sera91}\apj{Serabyn, E., Lacy, J.H.
\& Achtermann, J.M. 1991}{378}{557}
\bibitem[Telesco, et al. 1996]{tel96}\apj{Telesco, C.M., Davidson, J.A.
\& Werner, M.W. 1996}{456}{541}
\bibitem[Wolfire, Tielens \& Hollenback 1990]{wolf90}\apj{Wolfire, 
M.G., Tielens, A. \& Hollenbach, D. 1990}{358}{116}
\bibitem[Yusef-Zadeh \& Melia 1992]{ym92}\apjlett{Yusef-Zadeh, F. \& Melia, F. 1992}
{385}{41L}
\bibitem[Yusef-Zadeh \& Morris 1987]{ym87}\apj{Yusef-Zadeh, F. \&
Morris, M. 1987}{320}{545}
\bibitem[Yusef-Zadeh et al. 1997]{yz97}\refindent Yusef-Zadeh, F., Purcell,
W., \& Gotthelf, E. 1997, Proceedings of the
Fourth Compton Symposium, (New York: AIP), 1027.
\bigskip\bigskip
\end{thebibliography}
\end{document}